\documentclass[letter,oldversion]{aa} 
\usepackage{amsmath, amssymb}
\usepackage{subfigure} 
\usepackage{textcomp}
\usepackage{graphicx}
\usepackage{footmisc}
\usepackage[authoryear]{natbib}

\bibpunct{(}{)}{;}{a}{}{,}
\usepackage{txfonts}
\begin{document}
   \title{LABOCA and MAMBO-2 imaging of the dust ring of the Sombrero galaxy (NGC\,4594)
\thanks{This publication is 
based on data acquired with the IRAM 30m telescope and the
Atacama Pathfinder Experiment (APEX). APEX is a collaboration between the 
Max-Planck-Institut fur Radioastronomie, the European Southern Observatory, 
and the Onsala Space Observatory.}}

   \titlerunning{LABOCA and MAMBO-2 imaging of the dust ring of NGC\,4594}
  
   \author{C. Vlahakis
          \inst{1}
          \and
          M. Baes
          \inst{2}
          \and
          G. Bendo
          \inst{3}
          \and
          A. Lundgren
          \inst{4}          
          }

   \offprints{C. Vlahakis}

   \institute{Leiden Observatory, Leiden University, P.O. Box 9513, NL-2300 RA Leiden, The Netherlands\\
              \email{vlahakis@strw.leidenuniv.nl}
         \and Sterrenkundig Observatorium, Universiteit Gent, Krijgslaan 281 S9, 9000 Gent, Belgium
         \and Astrophysics Group, Imperial College, Blackett Laboratory, Prince Consort Road, London SW7 2AZ, UK 
         \and ESO, Casilla 19001, Santiago 19, Chile
             }

   \date{Received ; accepted }
 
  \abstract{The Sombrero galaxy (NGC\,4594) is an Sa galaxy with a symmetric dust ring. 
We have used the Large APEX BOlometer CAmera (LABOCA) at 870\hbox{\,\textmu m}\/ 
and the MAx-Planck Millimeter BOlometer (MAMBO-2) at 1.2\,mm to detect 
the dust ring for the first time at submillimetre and millimetre wavelengths. We have 
constructed a model of the galaxy to separate the active galactic nucleus (AGN) 
and dust ring components. The ring radius at both 870\hbox{\,\textmu m}\/ and 1.2\,mm agrees 
well with the radius determined from optical absorption and atomic gas studies. 
The spectral energy 
distribution of the ring is well fitted by a single grey-body with dust emissivity index 
$\beta=2$ and a dust temperature $T_{\mathrm d}$=18.4\,K. The dust mass of the 
ring is found to be $1.6\pm0.2\times 10^7$\,$M_\odot$ which, for a Galactic gas-to-dust ratio, 
implies a gas mass that is consistent with measurements from the literature.}


   \maketitle
%

\section{Introduction}

The Sombrero galaxy is one of the best studied early-type 
spiral galaxies. 
It is an Sa galaxy with a prominent dust lane, 
which is actually a symmetric dust ring 
seen nearly edge-on, and a nucleus that hosts a $10^9M_\odot$ 
supermassive black hole \citep{kormendy}.
These two features
are the highest surface brightness infrared sources in the galaxy 
\citetext{\citealt{bendo-a}, hereafter B06}.
At a distance of 9.4\,Mpc (the average of measurements from \citealt{ford} 
and \citealt{tonry}), it is an ideal 
laboratory for investigating different astrophysical processes, in particular, the 
connection between star formation and an active galactic nucleus (AGN) and the 
effects of these mechanisms on nuclear and global galaxy properties. 
The galaxy is also ideal for studying the separate spectral energy distributions (SEDs) of 
diffuse interstellar dust and the dust heated by an AGN, since the majority of 
diffuse dust is located in a ring relatively far from the AGN, 
making it thus relatively simple to model either component (B06).

Until recently, investigation of this galaxy in the far-infrared (FIR) and submillimetre 
(submm) \citep[e.g.][]{rice,krause} was limited to the global SED 
due to the resolution of \textit{IRAS}. B06 investigated the 
mid- and far-infrared SED of the separate components using \textit{Spitzer} 
(3.6--160\hbox{\,\textmu m}) and SCUBA (850\hbox{\,\textmu m}) 
observations. However, only the AGN component was detected with SCUBA, and the SCUBA 
map covered only $\sim3\arcmin \times 2\arcmin$, while the diameter 
of the dust ring as seen by \textit{Spitzer} measures $\sim$5\arcmin\/ (B06).

We present observations of the Sombrero galaxy (NGC\,4594) 
at 870\hbox{\,\textmu m}\/ with the Large APEX BOlometer CAmera 
\citep[LABOCA;][]{kreysa03,siringo} and at 1.2\,mm with the 
MAx-Planck Millimeter BOlometer \citep[MAMBO-2;][]{kreysa98}. 
In order to separately examine the dust ring and the AGN, we have 
modelled the dust emission using the method described in B06. 
We present the first results from this analysis, and focus on the FIR/submm 
emission from the dust ring.
Investigation of the FIR/submm emission from the nucleus will be the subject 
of a separate paper (Vlahakis et al. in prep.).

\section{Observations and Data Reduction}
\label{Observations}   

\begin{figure*}
\begin{center}
\includegraphics[angle=0,width=17.5cm,clip]{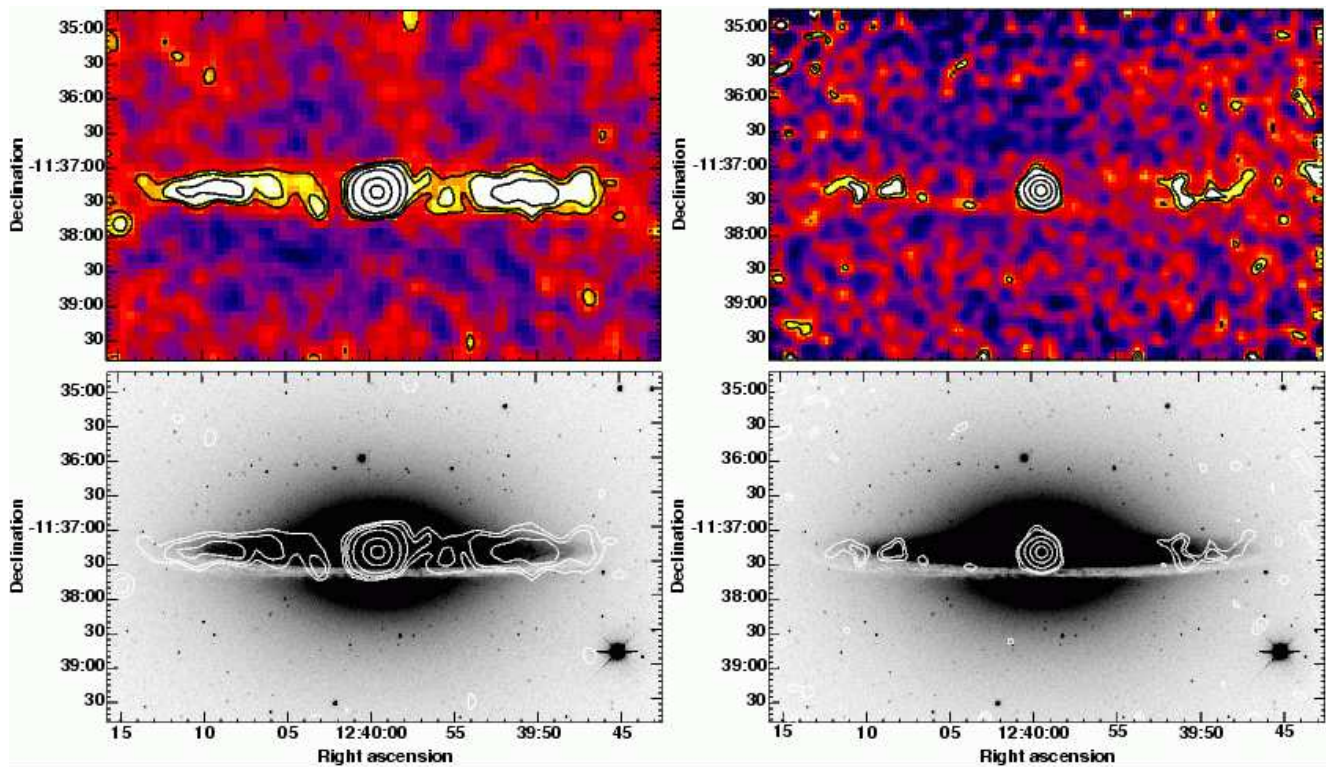}
\caption{\label{LABOCA+MAMBO-map}{Submm/mm images of the Sombrero galaxy. 
Left panel: 870\hbox{\,\textmu m}\/ LABOCA map, smoothed to a resolution of $\sim$22\arcsec, 
shown as a contoured colour scale (top) and as 
870\hbox{\,\textmu m}\/ contours overlaid onto an optical image (bottom). 
The rms in the smoothed map is 3.4~\mbox{mJy\,beam$^{-1}$}. 
Contours show flux densities of 6.8, 10.2, 17, 34, 85, and 136~\mbox{mJy\,beam$^{-1}$}. 
The central $8\arcmin \times 5\arcmin$ of the map is shown. 
Right panel: 1.2\,mm MAMBO-2 map, smoothed to a resolution of $\sim$15\arcsec, 
shown as a contoured colour scale (top) and as 
1.2\,mm contours overlaid onto an optical image (bottom). 
The rms in the smoothed map is 1.1~\mbox{mJy\,beam$^{-1}$}. 
Contours show flux densities of 2.2, 3.3, 11, 33, and 66~\mbox{mJy\,beam$^{-1}$}.  
The central $8\arcmin \times 5\arcmin$ of the map is shown; we note that the peaks at the edges of the 
image are noise features towards the edges of the map.}}
\end{center}
\end{figure*}

The 870\hbox{\,\textmu m}\/ data were taken with the newly commissioned LABOCA bolometer array, 
located on the Atacama Pathfinder EXperiment (APEX) telescope on Chajnantor, 
Chile, during its science verification phase in May 2007. The full width half maximum (FWHM) of the point 
spread function (PSF) at 870\hbox{\,\textmu m}\/ is $\sim$19\arcsec. The observations were carried out using 
a raster-spiral observing mode \citep{weiss} in which each scan consists of 4 spiral pointings offset 
by 27\arcsec, each spiral providing a fully sampled map of the full field of view 
(FOV) of LABOCA (11.4\arcmin). 
The total map thus covers an area slightly larger than the FOV. 
The total on-source integration time was 40 minutes, reaching an 
rms level of $\sim$8~\mbox{mJy\,beam$^{-1}$} in the unsmoothed map. 
We reduced the data with the BoA\footnote{http://www.astro.uni-bonn.de/boawiki/Boa} software package. 
The data reduction process included correction for 
atmospheric zenith opacity (opacity, $\tau$, values ranged from 0.2 to 0.3), 
flat-fielding, despiking, correlated sky noise removal, 
and the removal of further correlated noise due to instrumental effects.
The data were calibrated using observations of Mars and Uranus. 
The uncertainty due to calibration is the main source 
of uncertainty, and is estimated to be $\sim$10\%.
For presentation purposes, we smoothed the final map to a resolution 
of $\sim$22\arcsec\/ by convolving it with a Gaussian.

We carried out the 1.2\,mm observations with the MAMBO-2 117-element bolometer array 
on the 30m telescope on Pico Veleta, Spain, during the pooled observing 
sessions in December 2006 and March 2007. 
The FWHM of the PSF at 1.2\,mm is $\sim$11\arcsec. 
We used a standard on-the-fly mapping technique and a wobbler switching mode with a 
chop throw of 240\arcsec\/ to map an area of 
$\sim9\arcmin \times 5\arcmin$. The total integration time was 96 minutes, 
reaching an rms of $\sim$3~\mbox{mJy\,beam$^{-1}$} in the unsmoothed map. 
Opacity ($\tau$) values ranged from $<$0.1 to 0.3, 
while the sky noise was consistently low. We reduced the data using standard methods in MOPSIC, 
part of the GILDAS software package. 
This included correction for atmospheric opacity, flat-fielding, 
despiking, and removing correlated sky noise. 
The data were calibrated using observations of standard calibration sources, 
resulting in a calibration uncertainty of $\sim$15\%. For presentation purposes, we 
smoothed the final map to a resolution of $\sim$15\arcsec\/ by convolving it with a Gaussian.

\begin{figure*}
\begin{center}
\includegraphics[angle=0, width=17.5cm,clip]{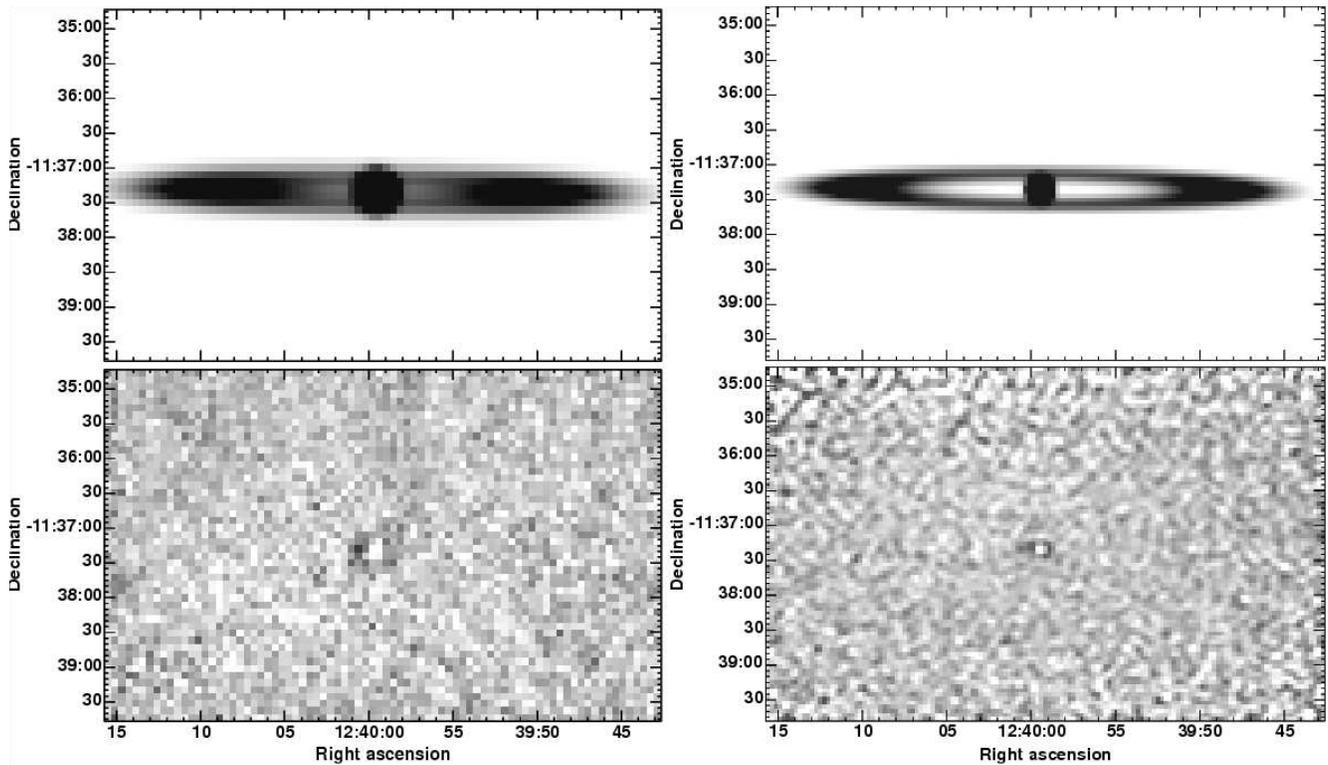}
\caption{\label{LABOCA+MAMBO-model}{Image model (top) and residual image (bottom) made by subtracting 
the model image from the observed image for the LABOCA 870\hbox{\,\textmu m}\/ data (left panel) 
and the MAMBO-2 1.2\,mm data (right panel).}}
\end{center}
\end{figure*}

\section{Results}
\label{Results}

The 870\hbox{\,\textmu m}\/ LABOCA image and the 1.2~mm MAMBO-2 image are shown in 
Fig.~\ref{LABOCA+MAMBO-map}. 
Contours showing levels of equal flux density are overlaid onto the images. 
These same contours are also shown overlaid onto an optical B-band image 
\citep[from the Spitzer Infrared Nearby Galaxies 
Survey (SINGS) Legacy Data;][]{kennicutt}. 
The dust ring and the nucleus are clearly detected in the LABOCA map. 
While the nucleus was detected previously at 
850\hbox{\,\textmu m}\/ with SCUBA and at 
870\hbox{\,\textmu m}\/ with the Heinrich Hertz Telescope 
(B06; \citealt{krause}, respectively), here the 
dust ring is detected for the first time at 
submm wavelengths. The dust ring and nucleus are also detected in the MAMBO-2 map. 
In both the submm and mm images, there is no evidence of dust 
outside the ring or outside the plane of the ring, and we do not detect any 
substructure in the ring at this signal-to-noise level.

In order to separate the nuclear and ring components, we fitted a model to the unsmoothed data 
using the method described in B06. 
To summarise this method, the model consists of two components, a Gaussian ring and a central 
point source, which were convolved with a Gaussian PSF
with a FWHM equivalent to that for LABOCA and MAMBO-2; the fit was performed over a region 
corresponding to the optical disk.
The image models are shown in Fig.~\ref{LABOCA+MAMBO-model} 
together with the residuals from the model fits.

The 870\hbox{\,\textmu m}\/ and 1.2\,mm flux densities for the two components are given in 
Table~\ref{Fluxes}, together with the global flux density. 
The uncertainties given are those resulting from the model fitting process, 
while the largest source of uncertainly comes from the calibration. We note that 
the 870\hbox{\,\textmu m}\/ flux density we find for the nucleus is in 
excellent agreement with previous submm observations \citetext{B06; \citealt{krause}}, 
which are listed in Table~\ref{Fluxes} for comparison. 
The ring model parameters are given in Table~\ref{ring-parameters}, along with the 
parameters determined from the 5.7--70\hbox{\,\textmu m}\/ data by B06.  

\begin{figure}
\begin{center}
\includegraphics[angle=270, width=8.8cm]{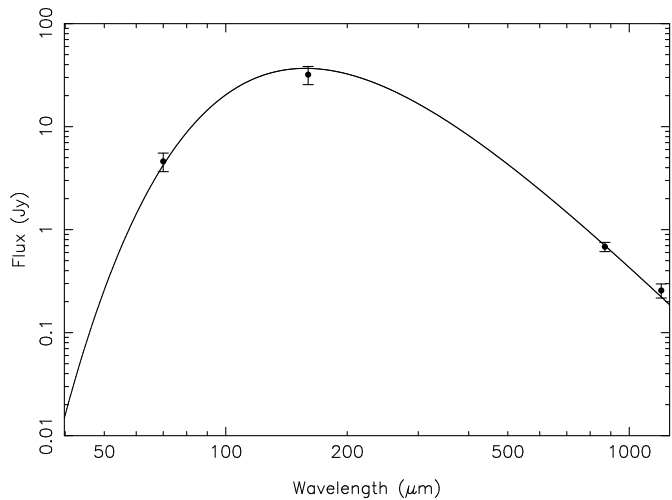}
\caption{\label{sed}SED of the dust ring. A single grey-body with $\beta$=2 is fitted to the 
70 and 160\hbox{\,\textmu m}\/ (\textit{Spitzer}), 870\hbox{\,\textmu m}\/ (LABOCA), and 1.2\,mm (MAMBO-2) 
flux densities. The fitted dust temperature is 18.4\,K.}
\end{center}
\end{figure}

We have fitted a modified Planck function (`grey-body') to the 70\hbox{\,\textmu m}--1.2\,mm SED of the dust ring, 
using the 70- and 160-\hbox{\,\textmu m}\/ \textit{Spitzer} measurements from B06 and our 
870-\hbox{\,\textmu m}\/ and 1.2\,mm measurements from Table~\ref{Fluxes}. 
In this fit, the Planck function was modified by a 
$\lambda^{-\beta}$ emissivity function with dust emissivity index $\beta$=2, and the dust temperature 
$T_{\mathrm d}$ was allowed to vary. The SED of the dust ring is found to be well 
fitted by a single grey-body with $T_{\mathrm d}$=18.4\,K (Fig.~\ref{sed}). 

\begin{table}
\caption{\label{Fluxes} Flux densities for the separate components of NGC\,4594.}
\begin{minipage}[t]{\columnwidth}
\centering
\renewcommand{\footnoterule}{} 
\begin{tabular}{lccc}
\hline\hline
                    & \multicolumn{3}{c}{Flux density (Jy)} \\
Wavelength          & \cline{1-3} 
(\hbox{\,\textmu m})           &  Global & Nucleus/AGN      & Ring \\ 
\hline
1200                &  0.442 & 0.185$\pm$0.002\footnote{Uncertainties are from the fits to the data and do not include calibration uncertainties (see Section~\ref{Observations}).\label{fn:uncert}}  & 0.257$\pm$0.006\\ 
870                 &  0.924 & 0.242$\pm$0.003\footref{fn:uncert}  & 0.682$\pm$0.016\\
870                 &  ...   & 0.230$\pm$0.035\footnote{Previous submm measurements \citetext{B06; \citealt{krause}}.\label{fn:comp}}     & ... \\
850                 &  ...   & 0.25$\pm$0.06\footref{fn:comp}    & ... \\
\hline
\end{tabular}
\end{minipage}
\end{table}

\begin{table}
\caption{\label{ring-parameters}Ring model parameters from fits to data.}
\begin{minipage}[t]{\columnwidth}
\centering
\renewcommand{\footnoterule}{} 
\begin{tabular}{lccc}
\hline\hline
Parameter\footnote{Definitions of the parameters are as given by B06.} & 1.2\,mm            & 870\hbox{\,\textmu m}                & 5.7--70\hbox{\,\textmu m}\footnote{Values from B06.}\\
\hline
Min/Maj axis ratio & 0.0757$\pm$0.0014          & 0.0710$\pm$0.0013           & 0.0990$\pm$0.0015\\
Position angle         & 90.43$^\circ\pm$0.04$^\circ$ & 90.69$^\circ\pm$0.09$^\circ$ & 89.25$^\circ\pm$0.06$^\circ$\\
Ring radius            & 169.9\arcsec$\pm$1.0\arcsec  & 162.0\arcsec$\pm$3.0\arcsec  & 144.7\arcsec$\pm$0.5\arcsec\\
Ring width             & 33.0\arcsec$\pm$2.0\arcsec   & 45.7\arcsec$\pm$0.6\arcsec    & 22.9\arcsec$\pm$1.1\arcsec\\
\hline
\end{tabular}
\end{minipage}
\end{table}

We estimate the dust mass of the ring with
\begin{equation} \label{eq:dmass}
M_{\mathrm d}=\frac{S_{\mathrm870} D^{2}}{\kappa_{\mathrm870}B(T_{\mathrm d})_{\mathrm870}}
\end{equation}
where $S_{\mathrm 870}$ is the 870\hbox{\,\textmu m}\/ flux density; D is the distance (9.4\,Mpc); 
$B(T_{\mathrm d})_{\mathrm 870}$ is the Planck function for the fitted temperature 
$T_{\mathrm d}$; and $\kappa_{\mathrm 870}$ is the dust
 mass opacity coefficient at 870\hbox{\,\textmu m}. 
Assuming a value $\kappa_{\mathrm 870}=0.041$\,m$^{2}$\,kg$^{-1}$ 
\citep{li-draine}, we obtain $M_{\mathrm d}=1.6\pm0.2\times 10^7$\,$M_\odot$. 
Here, the uncertainty is derived from the uncertainty on the 870\hbox{\,\textmu m} flux 
density, which is the main source of uncertainty for the measured parameters. 
We note, however, that a potentially larger source of uncertainty could arise due to the 
value of $\kappa$, which is notoriously uncertain.

\section{Discussion}
The dust ring detected at 870\hbox{\,\textmu m}\/ and at 1.2\,mm is wider than the 
dust ring detected at 5.7--70\hbox{\,\textmu m}. 
This indicates that the dust ring is not monochromatic; the dust on the inside of the 
ring is warmer and radiates primarily at shorter wavelengths than the dust on the outside.  
This interpretation is consistent with the dust temperature gradient suggested by the 
analysis of dust opacity measurements presented by \citet{emsellem}. 
These results imply that the primary heating sources for the ring are located on the inside 
of the ring.

The $\sim$170\arcsec\/ radius of the dust ring found at 870\hbox{\,\textmu m}\/ and 1.2\,mm is 
consistent with the outermost of two rings determined in optical absorption 
\citep[e.g.,][]{dettmar,wainscoat,emsellem} and neutral hydrogen (H{\sc{i}}) emission 
\citep{bajaja84}.
The 5.7--70\hbox{\,\textmu m}\/ dust, molecular gas, radio synchrotron emission, and part of the 
neutral hydrogen (H{\sc{i}}) emission lies in a $\sim$140\arcsec\/ ring \citep{bajaja88,bajaja91,krause}. 
This ring is consistent with the innermost of the two rings detected in optical absorption. 
Thus, it appears that emission from wavebands more commonly associated with star formation 
originates from the $\sim$140\arcsec\/ ring whereas `quiescent' dust and gas emission, 
not necessarily associated with star formation, originates from the $\sim$170\arcsec\/ ring.

The 1.2\,mm data point is useful for constraining the 
Rayleigh-Jeans (RJ) tail of the SED, yet we note that the 70-, 160-, and 870-\hbox{\,\textmu m}\/ data alone
are also well fitted by this grey-body. This result is in contrast with 
\citet{galliano} and \citet{bendo-b}, who found that the RJ side of
the SED deviates from thermal dust emission at wavelengths longer than 850\hbox{\,\textmu m}.

The dust mass we calculated from our measured 870\hbox{\,\textmu m}\/ flux density and fitted 
dust temperature is about a factor of two 
higher than the dust mass estimated from optical absorption by \citet{emsellem} 
(8.3$\times 10^6$\,$M_{\odot}$, scaled to 9.4\,Mpc). Part of this discrepancy can probably be 
explained if there is an inconsistency between the dust opacity model used by 
\citet{emsellem} and the \citet{li-draine} model.
Our estimated dust mass is also 
a factor of two higher than found by B06 using the FIR measurements alone. 
This is not surprising, since the RJ part of the SED, which is sampled at the LABOCA 
and MAMBO-2 wavelengths, is more sensitive to the mass of dust and less sensitive to temperature. 
\citet{draine} comment that the models
used in B06 to estimate dust masses are only accurate
to within a factor of 1.5 when no submm data are available. Thus, the 
additional submm/mm data points are needed to better trace the bulk of the dust mass. 

Adopting a Galactic gas-to-dust ratio \citep[165;][]{li}, our dust mass implies a total gas mass of 
2.6$\times 10^{9}$\,$M_{\odot}$. 
We can compare this value to the total (molecular + atomic) 
gas mass using measurements from the literature. \citet{bajaja84} find an H{\sc{i}} 
mass of 3.3$\times 10^{8}$\,$M_{\odot}$ (scaled to our adopted distance of 9.4\,Mpc). 
Preliminary results from \citet{sengupta} are in good agreement with this value. 
\citet{bajaja91} used their detection of CO at one position in the ring to estimate an upper limit 
to the molecular gas mass of 4.6$\times 10^{8}$\,$M_{\odot}$ (scaled to 9.4\,Mpc), though they note that 
this value could be uncertain by up to a factor of five. 
Thus, the total gas mass implied by our dust mass is consistent with measurements 
of the total (molecular + atomic) gas mass.

\section{Conclusions}

We have observed the Sombrero galaxy (NGC\,4594) at 870\hbox{\,\textmu m}\/ 
with the LABOCA bolometer camera at the APEX telescope and at 1.2\,mm with 
the MAMBO-2 bolometer array at the IRAM 30m telescope.

We have detected the dust ring for the first time at submm 
and mm wavelengths. We have constructed a model of the galaxy to separate the 
AGN and dust ring components. The ring radius at both 870\hbox{\,\textmu m}\/ and 1.2\,mm agrees 
well with the radius determined from optical absorption and atomic gas studies.

We find that the 70\hbox{\,\textmu m}--1.2\,mm SED of the dust ring is well fitted  
by a single grey-body with $\beta$=2 and a dust 
temperature $T_{\mathrm d}$=18.4\,K. Using this dust temperature we find the ring to 
have a dust mass of  $1.6\pm0.2\times 10^7$\,$M_\odot$.
For a Galactic gas-to-dust ratio, our estimated dust mass implies a gas mass 
that is consistent with measurements from the literature.

\begin{acknowledgements}
Thanks go to Alexandre Beelen, Frederic Schuller, and Axel Weiss for useful discussions 
about BoA data reduction, and to the anonymous referee for helpful comments. 
We are also grateful to the staff at the IRAM 30m telescope 
and the APEX telescope for their support with the observations.
\end{acknowledgements}

\bibliographystyle{aa}

\end{document}